\documentclass[conference]{IEEEtran}
\ifCLASSINFOpdf
\else
\fi
\usepackage{amsmath,amssymb,amsfonts}
\usepackage{graphicx}
\usepackage{textcomp}
\usepackage{xcolor}
\usepackage{comment}
\usepackage{float}
\usepackage{booktabs}
\usepackage{subfig}
\usepackage{amsmath}
\usepackage{adjustbox}
\usepackage{array}
\usepackage{booktabs}
\newcolumntype{L}{>{\centering\arraybackslash}m{0.4in}}

\usepackage{algorithm2e}
\usepackage{algpseudocode}
\RestyleAlgo{ruled}
\usepackage{graphicx}    % For including images
\usepackage{caption}     % For better control of captions (optional but helpful)
\usepackage{dblfloatfix} % Ensures figure* floats appear at the top of the page
\usepackage{enumitem} % Needed for custom list settings

\def\BibTeX{{\rm B\kern-.05em{\sc i\kern-.025em b}\kern-.08em
    T\kern-.1667em\lower.7ex\hbox{E}\kern-.125emX}}

\newcommand{\newchanges}[1]{{\color{black}  #1}}

\usepackage{setspace}
\usepackage{url}
\usepackage{hyperref}

\begin{document}
\title{FuSeFL: Fully Secure and Scalable Federated Learning}

% % affiliations
\author{\IEEEauthorblockN{Sahar Ghoflsaz Ghinani, Elaheh Sadredini}
	\IEEEauthorblockA{Department of Computer Science and Engineering\\
        University of California, Riverside\\
		\{sghof001, elahehs\}@ucr.edu}}
\maketitle

\begin{abstract}
Federated Learning (FL) enables collaborative model training without centralizing client data, making it attractive for privacy-sensitive domains. While existing approaches employ cryptographic techniques such as homomorphic encryption, differential privacy, or secure multiparty computation to mitigate inference attacks, including model inversion, membership inference, and gradient leakage, they often suffer from high computational and memory overheads. Moreover, many methods overlook the confidentiality of the global model itself, which may be proprietary and sensitive. 
These challenges limit the practicality of secure FL, especially in settings that involve large datasets and strict compliance requirements.

We present \textbf{FuSeFL}, a \textbf{Fu}lly \textbf{Se}cure and scalable \textbf{FL} scheme, which decentralizes training across client pairs using lightweight MPC, while confining the server's role to secure aggregation, client pairing, and routing.
This design eliminates server bottlenecks, avoids full data offloading, and preserves full confidentiality of data, model, and updates throughout training.
Based on our experiment, FuSeFL defends against unauthorized observation, reconstruction attacks, and inference attacks such as gradient leakage, membership inference, and inversion attacks, while achieving up to $13 \times$ speedup in training time and 50\% lower server memory usage compared to our baseline. 
\end{abstract}


% \fi
\IEEEpeerreviewmaketitle

\section{Introduction}

Federated Learning (FL) is a distributed machine learning paradigm that enables multiple clients to collaboratively train a global model without sharing their local data~\cite{bonawitz2019federatedlearningscaledesign,mcmahan2023communicationefficientlearningdeepnetworks,8241854,10.1145/3133956.3133982,McMahan2017,mcmahan2018,prio}. This paradigm is particularly suited for scenarios where legal, privacy, or infrastructural constraints prevent centralized data collection~\cite{kairouz2021advancesopenproblemsfederated,McMahan2017}. In FL, a central server distributes a model to the clients, who perform local training and return updates, such as gradients or model weights. These updates are aggregated to refine the global model, all while the raw data remains on client devices, preserving confidentiality and reducing transmission costs.
While FL offers clear advantages in preserving data locality, it remains vulnerable to significant privacy breaches, where (1) submitted gradients or model updates can unintentionally encode patterns from the underlying data, allowing adversaries to extract sensitive information \cite{leak1,10.1145/3133956.3133982,rofl,elsa}, and (2) attackers may attempt to steal or misuse the offloaded model for malicious or competitive gain.

To mitigate \textbf{users' data privacy concerns}, a wide range of techniques have been proposed to protect the privacy of clients’ data during training. 
\emph{Homomorphic encryption} (HE) enables secure aggregation over encrypted updates without decryption~\cite{8241854,10.1145/3133956.3133982}, but incurs high computational and memory costs, often slowing down training by several orders of magnitude~\cite{survay}. \emph{Differential privacy} (DP) defends against inference attacks by adding noise to updates~\cite{McMahan2017,mcmahan2018}, though stronger privacy budgets degrade model accuracy~\cite{survay}. \emph{Secure multi-party computation} (MPC) allows collaborative aggregation without exposing individual inputs~\cite{prio,prioplus,elsa,ppfl, 10.1145/3133956.3133982,291112}, but communication overhead grows with the number of clients~\cite{survay}.

Beyond data privacy, a critical yet often underappreciated concern in FL is the \textbf{privacy of the model itself}, particularly from the perspective of the model owner. In typical FL deployments, the global model, often developed through significant intellectual and financial investment, is shared with all participating clients, some of whom may be semi-honest or untrusted~\cite{tramèr2016stealingmachinelearningmodels}. This exposes the model to risks such as extraction, reverse engineering, or unauthorized use. The concern becomes even more pressing when clients are typically large institutions, such as hospitals, banks, or research labs, that operate under strict privacy regulations and contractual obligations. In such settings, the model is not only a valuable proprietary asset but also a potential vector for secondary leakage of sensitive information. For example, attackers may perform \emph{model inversion} to reconstruct private inputs or launch \emph{membership inference} attacks to determine whether a particular data was part of the training set~\cite{10.1145/2810103.2813677}. Ensuring model confidentiality is therefore essential, not only to protect intellectual property but also to comply with legal, regulatory, and contractual boundaries. Several recent works have attempted to address model privacy in FL\cite{wwfl,secureml,securenn,falcon,ryffel2021ariannlowinteractionprivacypreservingdeep,fanfour,fssnn}, though each approach faces significant limitations that hinder its practicality, particularly in settings where efficiency, scalability, and regulatory compliance are critical.
Watermarking-based techniques have been proposed to address model ownership and intellectual property protection~\cite{waffle, liang2023fedcipfederatedclientintellectual}. These methods embed verifiable patterns, such as trigger inputs or parameter-level watermarks, into the model, allowing the owner to later demonstrate unauthorized use. However, watermarking does not prevent theft or misuse during training; it only enables detection after the fact. In domains like healthcare and finance, where compliance and security must be enforced upfront, such reactive mechanisms are insufficient.

 WW-FL~\cite{wwfl} adopts a fully offloaded training architecture in which MPC is performed across distributed clusters of servers. While this design offers strong security guarantees, it requires clients to upload their entire datasets to third-party servers, contradicting the foundational principle of FL, where data should remain local to each client. In addition, this architecture introduces a trade-off between communication overhead, memory usage, and the number of server clusters. When dealing with thousands of clients, increasing the number of servers raises the system's cost while reducing communication time and memory usage. 
 
Moreover, WW-FL depends on geographically distributed compute infrastructure, which elevates deployment complexity and operational costs. In addition, in settings such as cross-silo environments, where clients (hospitals or financial institutions) already possess rich compute and storage capabilities, outsourcing training to external servers is not only inefficient but also unnecessary.

A similar line of work proposes offloading both the model and client data to centralized secure servers that perform training using cryptographic techniques \cite{flsurvey} such as MPC~\cite{secureml,securenn,falcon,ryffel2021ariannlowinteractionprivacypreservingdeep,fanfour,fssnn}. While these systems reduce the trust placed in individual clients, they impose substantial communication burdens due to repeated transfers of large training datasets. 
Additionally, centralizing sensitive datasets on untrusted servers risks violating regulatory constraints, institutional data policies, and contractual agreements, making these approaches ill-suited for regulated domains.

AriaNN~\cite{ryffel2021ariannlowinteractionprivacypreservingdeep} introduces a hybrid FL scheme that jointly protects model and data privacy by secret-sharing the model between the clients and server using MPC. This design prevents either party from observing the full model or training data in plaintext. While more aligned with FL principles, AriaNN suffers from serious scalability limitations. Each client must train its share of the model in coordination with the server, typically in a sequential fashion due to server-side compute constraints. 
Furthermore, the server must maintain a dedicated model instance per client in addition to the global model, leading to unsustainable memory overhead for large-scale models such as VGG-16.
In summary, existing fully secure FL approaches struggle to scale beyond a few tens of clients due to excessive server-side computational and memory overheads. Techniques like HE and secure model sharing, while effective in preserving privacy, impose severe performance costs. In particular, efforts to simultaneously protect both data and model confidentiality exacerbate these challenges, rendering current solutions impractical for real-world, large-scale deployments.

To address the limitations of prior work, we propose \emph{FuSeFL}, a fully secure and scalable federated learning (FL) framework. FuSeFL ensures end-to-end confidentiality of both data and models while efficiently scaling to tens of thousands of clients without overloading the server. Its core idea is to decentralize training across dynamically formed client groups, while confining the server’s role to lightweight secure aggregation, client pairing, and data routing. In FuSeFL, clients are dynamically paired to form collaborative groups, where the global model is secret-shared between the two clients, and training is jointly executed using MPC over their secret-shared local data. This setup prevents any participant from accessing data or model parameters in plaintext. After local training, each client pair submits secret-shared updates to two servers, which perform secure aggregation without learning any private information.

To prevent collusion and leakage, FuSeFL assumes that at least one of the two aggregation servers is trusted and thus treated as non-colluding. This trusted server acts as the pairing authority, using available metadata to minimize the likelihood of repeatedly matching the same untrusted clients in different rounds. It also serves as a secure distribution mix, anonymously routing models and data between clients so that each remains hidden from its partner, effectively preventing intra-group collusion. This design eliminates server-side bottlenecks, avoids data offloading, and reduces memory overhead. Moreover, by secret sharing of both the model and the updates throughout training and aggregation, FuSeFL provides robust protection against gradient leakage, model update inference, membership inference, and model inversion attacks. Our evaluation shows that FuSeFL (1) reduces server memory usage by nearly 50\% compared to AriaNN-FL, (2) maintains nearly constant computation time as the number of clients increases, achieving linear throughput scalability and up to \textbf{13$\times$} and \textbf{9$\times$} speedups compared to AriaNN-FL and WW-FL, respectively. and (3) improves model accuracy by up to \textbf{1.1\%} over AriaNN-FL on MNIST.

\vspace{0.1cm}

\textbf{Our key contributions are:}
\begin{itemize}[nosep]
    \item \textbf{A fully secure FL scheme:} We develop \emph{FuSeFL}, a FL scheme that holistically secures model parameters, client data, and intermediate updates using lightweight MPC across client groups, addressing both model and data privacy within a unified design (\S\ref{intro}).

    \item \textbf{Decentralized and scalable training:} FuSeFL eliminates server-side training bottlenecks by distributing heavy computations among client pairs, and it only relies on the server for performing lightweight computations (\S\ref{sec:computation}).  

    \item \textbf{Practical efficiency with robust privacy guarantees:} Our approach significantly reduces memory and computation overheads while mitigating key inference threats such as model inversion, gradient leakage, and membership inference, making FuSeFL scalable to real-world deployments involving many clients (\S\ref{sec:memory}, \S\ref{sec:computation}).
\end{itemize}

% \vspace{-3mm}
\section{Motivation and Goal}

\subsection{Limitations of Existing Work} 

\begin{figure}[htp]
\centerline{\includegraphics[width=3in, trim={0cm 0.3cm 0cm 0.5cm}]{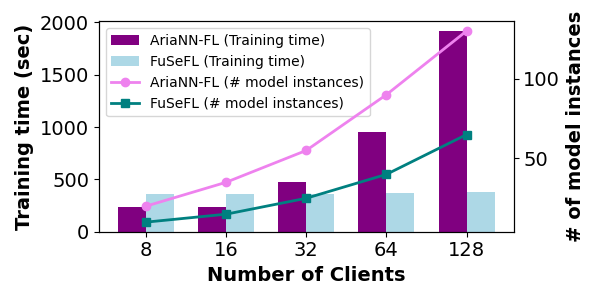}}
\caption{AriaNN-FL vs. FuSeFL, (Left vertical axis) in terms of secure training time for one epoch and (Right vertical axis) in terms of the number of model instances stored in one aggregator. }

\label{introfig}
\end{figure}

Despite increased attention to privacy in FL, fully secure FL remains constrained by fundamental system bottlenecks. We identify three key limitations:

\begin{enumerate}[nosep,leftmargin=*]
    \item \textbf{Model Exposure and Trust Assumptions:} Many FL schemes expose the global model to clients during training, risking model theft or misuse, particularly in regulated domains where model is sensitive \cite{im2,waffle,liang2023fedcipfederatedclientintellectual}. Moreover, several privacy-preserving methods, such as WW-FL\cite{wwfl}, offload client data or models to trusted third-party servers using MPC or HE protocols~\cite{secureml,securenn,falcon,ryffel2021ariannlowinteractionprivacypreservingdeep,he1,he2} while assuming that the computing parties do not collude. These trust assumptions are not realistic or violate the core principle of FL, which requires data to remain within clients' boundaries \cite{kairouz2021advancesopenproblemsfederated,cross3}.

    \item \textbf{Server Bottleneck and Poor Scalability:} schemes such as AriaNN-FL\cite{ryffel2021ariannlowinteractionprivacypreservingdeep} impose significant computational load on the server. Secure training requires the server to engage individually with each client, saturating its compute resources and limiting parallelism. As shown in Fig.~\ref{introfig} (left axis), training time increases sharply as the number of clients grows, even when using a 16-core server. This sequential interaction model hinders scalability in larger deployments.

    \item \textbf{Storage Overheads:}
    Secure training introduces high memory demands. This is because the server must retain a copy of each client's model update until aggregation is complete. For large models like VGG-16, this results in excessive memory usage: as illustrated in Fig.~\ref{introfig} (right axis), AriaNN-FL stores 128 full model instances when training with 128 clients.

\end{enumerate}

% \vspace{-4mm}
\subsection{Our Goals}
FuSeFL aims to deliver a practical, fully secure, and scalable FL framework that safeguards both user data and the model itself, an often-overlooked dimension in secure FL literature. Designed for real-world deployments with potentially several thousands of clients, FuSeFL balances strict privacy guarantees with system-level scalability and efficiency. Specifically, FuSeFL seeks to:

\begin{itemize}[nosep,leftmargin=*]
    \item \textit{Ensure end-to-end confidentiality of data and models:} Unlike conventional FL, where clients have full access to the global model, FuSeFL guarantees that the model and its updates remain secret-shared and are never revealed in plaintext to untrusted parties.

    \item \textit{Remove server-side computational bottlenecks:} By decentralizing local training to client groups, FuSeFL avoids per-client training load imposed on servers in frameworks like AriaNN-FL. Servers' computations are instead limited to lightweight secure aggregation and client pairing, improving scalability and efficiency.

    \item \textit{Reduce memory overhead:} FuSeFL forms client pairs that collaboratively train and send a single secret-shared model update per group, rather than having each client transmit updates independently. This design nearly halves server-side memory usage, improving resource efficiency in large-scale deployments.

    \item \textit{Achieve secure scalability to a larger number of clients:} Training is parallelized across multiple client groups, allowing FuSeFL to maintain high throughput and low latency as the number of clients grows, overcoming the scalability limitations of existing secure FL protocols.
\end{itemize}

\section{Threat Model and Security Considerations} \label{thretmod}

\textbf{Deployment Model and Trust Assumptions:} FuSeFL targets scalable FL where each client, such as a hospital, financial institution, or mobile device, holds a distinct set of samples over a shared feature space, i.e., horizontal FL \cite{gargary2024systematicreviewfederatedgenerative}.  Clients are governed by regulatory frameworks and formal agreements such as HIPAA or GDPR, which limit the risk of deviation from the protocol.

The trust model assumes: (1) \textbf{Clients} do not know whom they are pairing with, as their communication is routed through a trusted party. \newchanges{Due to this anonymity, clients in the same group cannot collude with each other or with the servers. Collusion among clients outside each training pair is not problematic, since the shares they hold do not allow the reconstruction of sensitive information.} Clients are semi-honest, meaning that while they may attempt to infer private information, they still adhere to the protocol. (2) At least one of the \textbf{aggregation servers} is trusted, while the other is honest-but-curious. They do not collude with one another or with clients, and together they jointly perform MPC-based aggregation. (3) \textbf{The model owner} is trusted to manage protocol coordination but cannot access client data or intermediate model states.
To avoid collusion and information leakage, the trusted server is responsible for pairing clients using the available metadata and the risk-aware dynamic client grouping algorithm. All communication between clients in a group passes through the trusted server to maintain anonymity. The risk-aware dynamic client grouping algorithm further prevents repeated interactions by reassigning clients into new training groups in each round.

\begin{figure}[htp]
\centerline{\includegraphics[width=2.5in, trim={1cm 0cm 0.1cm 0cm}]{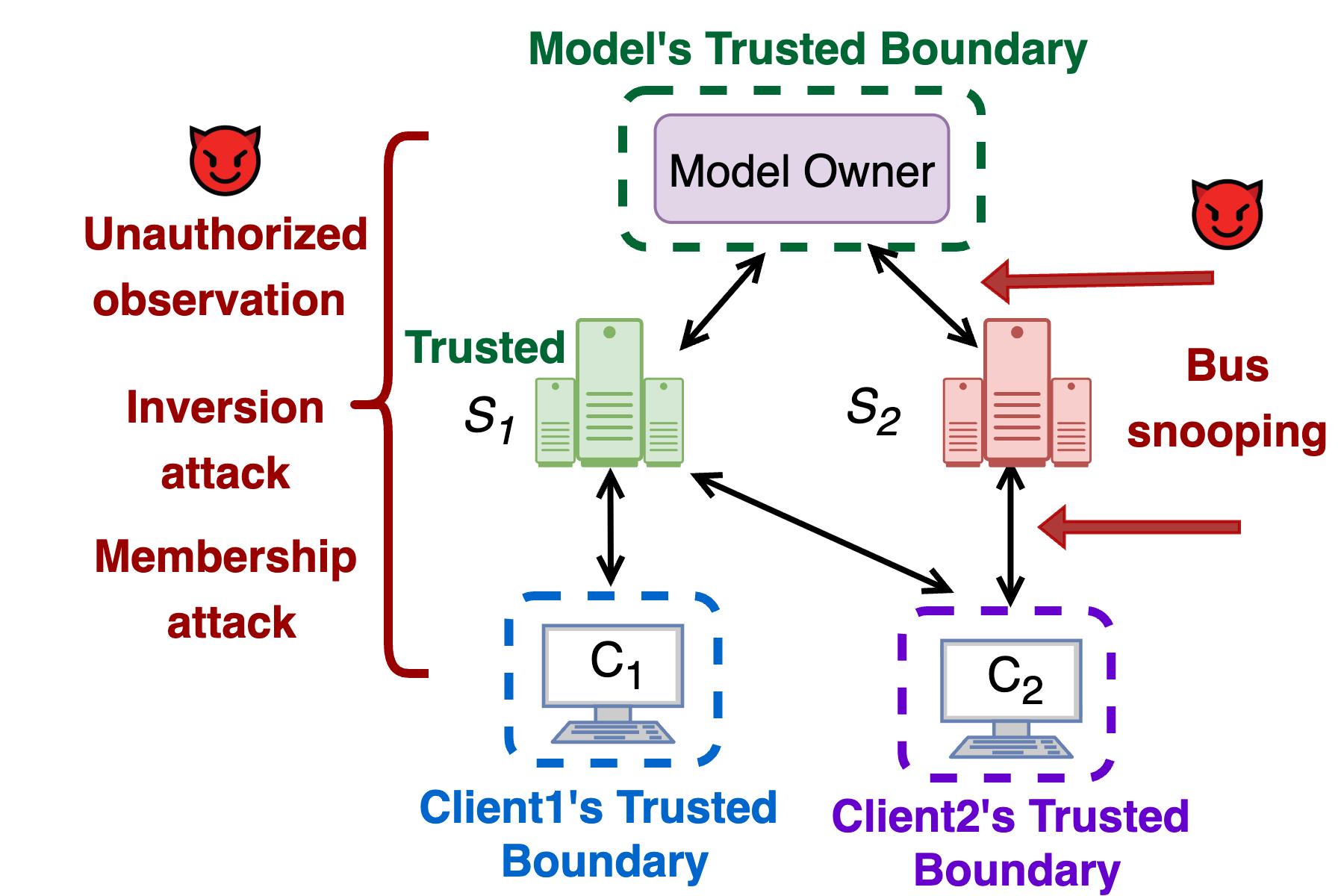 }}
\caption{Overview of the threat model. Each party retains plaintext within its trusted boundary. 
Adversaries attempt to infer private information outside the untrusted boundary.
}
\label{threat}
% \vspace{-0.2cm}
\end{figure}

\textbf{Threat Model:} We consider a \emph{semi-honest} (honest-but-curious) adversary who follows the protocol but may attempt to infer private information from observed data. The system involves three types of participants: clients, aggregation servers, and a model owner. Clients perform local training using secret-shared model parameters. Aggregation servers coordinate secure aggregation without learning the underlying values. The model owner is responsible for initializing the global model and orchestrating training (Fig. \ref{threat}).

\textbf{Security Guarantees:}
FuSeFL provides strong privacy protections through lightweight cryptographic primitives. Specifically:  
(1) \emph{Data confidentiality:} Each client's private data is secret-shared within its training group, ensuring that no single party can reconstruct the input.  
(2) \emph{Model confidentiality:} The global model is secret-shared between aggregation servers and is never exposed in plaintext to any party during training or aggregation.  
(3) \emph{Update privacy:} Local model updates are secret-shared and aggregated securely using MPC, such that individual updates remain hidden.

These guarantees mitigate risks such as unauthorized observation, gradient leakage, and reconstruction attacks. Since the model is never exposed in plaintext outside the trusted boundary, no untrusted party can access it to carry out membership inference or model inversion attacks aimed at extracting information about the training user data. 

\textbf{Out-of-Scope Threats:} FuSeFL does not currently defend against poisoning attacks \cite{vote} and malicious participants who deviate from the protocol, submit corrupted updates, or act in a Byzantine manner. Future extensions may incorporate robust aggregation or verifiable computation to address these threats. Side-channel attacks\cite{powtim,elc,traffic} such as timing\cite{powtim} or traffic analysis\cite{traffic} are out of scope and would require separate system-level countermeasures.

\section{Background}
\subsection{Notations and Preliminaries}

Our scheme leverages MPC over the ring~$\mathbb{Z}_{2^n}$ to ensure the confidentiality of training data and model throughout the computation. Table~\ref{tab:notations} provides a summary of the key notations used in the remainder of the paper.

\begin{table}[htp]

	\centering
 % \color{blue}
	\caption{List of Notations}
    % \vspace{-4mm}
	\scalebox{0.7}{
	\begin{tabular}{>{\centering\arraybackslash}p{1.3in}>{\centering\arraybackslash}p{3.1in}}
 % >{\centering\arraybackslash}p{0.9in}cccc}

		\toprule
  Notations&Explanation\\
		\midrule
  
    {$GM$}& {Global model in plaintext}\vspace{1mm}\\
    % {$[GM]_j$} & $j^{th}$ share of the secret shared global model \vspace{1mm}\\
    % $C$&  Vector ciphertext of size \textit{n} \vspace{1mm}\\
    {$LM_i$}&$i^{th}$ local model in plaintext\vspace{1mm}\\
    % $R$&  One-time Pads (OTPSs) of size \textit{n} \vspace{1mm}\\
    % {$[LM_i]_j$} &  $j^{th}$ share of the secret shared $i^{th}$ local model\vspace{1mm}\\
    {$D_i$}&Plaintext data of the $i^{th}$ client\vspace{1mm}\\
    % $R$&  One-time Pads (OTPSs) of size \textit{n} \vspace{1mm}\\
    {$[GM]_j$, $[LM_i]_j$, $[D_i]_j$} & $j^{th}$ share of the secret shared data/model of $i^{th}$ client\vspace{1mm}\\
    
		\bottomrule
	\end{tabular}}
	\label{tab:notations}
    % \vspace{-5mm}
\end{table}

\subsection{Federated Learning (FL)}

\begin{figure}[htp]
\centerline{\includegraphics[width=3.4in, trim={0cm 0cm 0cm 0.2cm}]{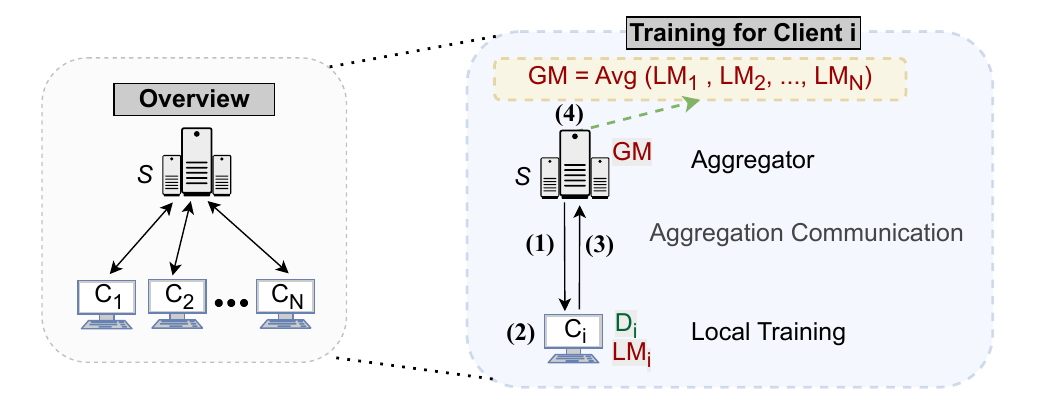}}
\caption{FL overview. (1) The server sends the global model to the clients. (2) The clients then train the model locally and (3) send back their trained models to the server. (4) The server updates the global model for the next round of training.}
\label{fl}
% \vspace{-0.2cm}
\end{figure}

Traditional machine learning requires aggregating data from multiple sources onto a central server for training. However, this approach is often impractical due to privacy regulations, communication costs, and bandwidth constraints~\cite{mcmahan2023communicationefficientlearningdeepnetworks}. 
FL addresses these challenges by enabling collaborative model training across distributed clients without exposing raw data~\cite{bonawitz2019federatedlearningscaledesign, mcmahan2023communicationefficientlearningdeepnetworks}. As illustrated in Fig.~\ref{fl}, (1) the central server first distributes the current global model to a subset of participating clients. (2) Each client trains the model locally using its private dataset. (3) The clients then send their model updates back to the server. (4) The server aggregates these updates (e.g., via FedAvg \cite{mcmahan2023communicationefficientlearningdeepnetworks}) to refine the global model. This process is repeated for multiple rounds until convergence.

FedAvg~\cite{mcmahan2023communicationefficientlearningdeepnetworks} is the most widely adopted aggregation method in FL. It computes the global model by taking a weighted average of the local model parameters ($w$) from participating clients:
\[
w_{t+1} = \sum_{k=1}^{K} \frac{n_k}{n} w_t^k
\]
Here, $w_t^k$ denotes the model parameters from client $k$ at round $t$, $n_k$ is the number of training samples on client $k$, and $n = \sum_{k=1}^{K} n_k$ is the total number of samples across all clients. This weighted averaging ensures that clients with larger datasets have proportionally greater influence on the updated global model. In FL, each client's dataset and the model size can amplify communication and memory demands, leading to scalability bottlenecks, particularly at the aggregator, where per-client updates must be processed and stored~\cite{flagger, bonawitz2019federatedlearningscaledesign, kairouz2021advancesopenproblemsfederated}. This occurs because we must keep a copy of all submitted updates to determine the weight associated with each client before performing the aggregation.

\subsection{Mix Networks (mixnets)}
Mix networks (mixnets) enable privacy-preserving communication by routing messages through a sequence of intermediate servers that shuffle, re-encrypt, and delay traffic to obscure the link between senders and receivers. This process provides strong anonymity by disrupting traffic analysis and timing correlations, effectively concealing communication patterns from adversaries \cite{mixnet1,mixnet2}. Although mixnets introduce additional latency due to multi-hop routing and layered encryption, modern low-latency designs (e.g., Loopix \cite{loopix}, Stadium \cite{stadium}) significantly mitigate this overhead, achieving delays of only a few seconds while preserving robust anonymity. Consequently, mixnets remain practical for privacy-sensitive applications such as anonymous email systems \cite{aemail1,aemail2} and electronic voting protocols \cite{evote1,evote2}.

\subsection{Secure Multi-Party Computation (MPC)} \label{back:mpc}

MPC is a cryptographic technique that enables multiple parties to collaboratively compute a function over their private inputs while ensuring that the inputs remain confidential and individual shares are never disclosed during the process. By splitting sensitive data into encrypted or secret shares and processing them separately, secure computation is achieved.
MPC allows computation over encrypted data while remaining lightweight, making it a practical option for distributing computation securely across multiple clients without incurring heavy computational costs. However, as the number of participants increases, the communication overhead in MPC protocols also grows significantly.

\textbf{Additive Sharing.}  
To perform linear operations, additive sharing~\cite{10.1007/978-3-642-32009-5_38,aby,ABY3,mpcdf} is a common MPC technique that splits private data into multiple secret shares, each distributed to a different party. In this approach, the data is secret-shared such that no single party can reconstruct it independently. We define additive secret sharing over a finite ring $\mathbb{Z}_{2^n}$, where $n$ is the bit-length of encoded values and arithmetic is performed modulo $2^n$.
Let $X \in \mathbb{Z}_{2^n}$ be a private value. In a setting with j parties, $X$ is split into j additive shares ($[x]_j$) and is represented as the sum of its shares:
% \vspace{-3mm}
\[
X = \sum [x]_j \mod{2^n}
\]
% \vspace{-1mm}
These shares are indistinguishable from random values and reveal no private information individually. Each party keeps its share private and does not communicate it to others. Secure computation is achieved by applying operations directly over these shares, as described in prior work~\cite{secureml,aby}.

\begin{figure}[htp]
\centerline{\includegraphics[width=2.8in, trim={0cm 0.3cm 0cm 0.5cm}]{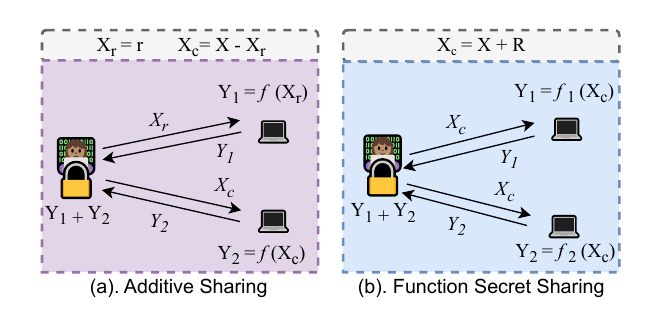}}
\caption{Examples of MPC schemes. (a) Additive sharing splits data into random shares for secure multi-party computation. (b) Function secret sharing splits the function itself across parties for privacy-preserving evaluation.
}
\label{smpc}
% \vspace{-0.4cm}
\end{figure}

\begin{itemize}[nosep,leftmargin=0pt, label={}, itemsep=0pt]
    \item[]
    \textit{Addition:} Each party performs addition locally on its share of the data. The partial results from all parties are then aggregated to compute the final result. As illustrated in Fig.~\ref{smpc}(a), the private input $X$ is split into two shares, $X_r$ and $X_c$. After local computation, the final output $Y$ is obtained by summing the partial results $Y_1$ and $Y_2$. For $j$ parties:
    % \vspace{-2mm}
    \[
    [Z]_j = [X]_j + [Y]_j \mod{2^n}
    \]
    % \vspace{-3mm}
    \[
    \quad Z = \sum \text{(locally computed)}\ [Z]_j \mod{2^n}
    \]
    \item[]
    \textit{Multiplication:}  
    To perform multiplication over additive shares, Beaver triples~\cite{10.1007/3-540-46766-1_34} are used. If $[Z] = [X] \cdot [Y]$, and the precomputed triple is $[c] = [a] \cdot [b]$~\cite{secureml}, each party $P_j$ does the following:
    
    \begin{itemize}[nosep]
        \item Compute local differences:
        % \vspace{-1mm}
        \[
        E_j = X_j - a_j, \quad F_j = Y_j - b_j
        \]
        % \vspace{-1mm}
        \item Publicly reconstruct the differences:
        % \vspace{-1mm}
        \[
        E = \sum E_j \mod{2^n}, \quad F = \sum F_j \mod{2^n}
        \]
        \item Compute the local share of the final result:
        % \vspace{-1mm}
        \[
        Z_j = j \cdot E \cdot F + F \cdot a_j + E \cdot b_j + c_j \mod{2^n}
        \]
    \end{itemize}
    
\end{itemize}

\textbf{Function Secret Sharing (FSS).}  
Unlike additive sharing, Function Secret Sharing~\cite{fss1,fss3,fss2} splits the function $f : \mathbb{Z}_{2^n} \rightarrow \mathbb{Z}_{2^n}$ across parties. Each party holds a share of the function rather than the data, and computation is done by applying these function shares to the input.
If $f(x)$ is the function to evaluate and $x$ is the (possibly masked) input, then:
% \vspace{-1mm}
\[
f(x) = f_1(x) + f_2(x) \mod{2^n}
\]
Fig.~\ref{smpc}(b) illustrates that the private input $X$ is masked as $X_c = X + R$, and each party evaluates its function share on $X_c$. The result is reconstructed by summing the outputs.
% \[
% f(X_c) = f_1(X_c) + f_2(X_c)
% \]

% \vspace{-0.3cm}

% \vspace{-2mm}
\subsection{AriaNN: Federated Learning Extension}\label{ariann}

Ryffel et al. \cite{ryffel2021ariannlowinteractionprivacypreservingdeep} proposed AriaNN, which uses additive sharing and function secret sharing (FSS) \cite{fss1,fss2} to privately perform deep learning tasks (both training and inference) with minimal communication overhead. In their scheme, both the data and the model are secret-shared between two parties to ensure that no single party can reconstruct the sensitive information. AriaNN supports several operations required for neural network training and inference.
AriaNN uses additive secret sharing to securely perform linear computations (explained in \S \ref{back:mpc}). For example, to support secure multiplication and convolution, AriaNN employs Beaver triples \cite{10.1007/3-540-46766-1_34}, specifically matrix-based Beaver triples \cite{secureml}, for efficient batch operations.
To perform comparison-based computations like ReLU, AriaNN uses FSS \cite{fss1,fss2} (explained in \S \ref{back:mpc}). By using FSS, AriaNN achieves a significant reduction in the number of communication rounds compared to previous works \cite{ABY3,securenn,falcon}. To implement comparison $(a \leq \alpha)$, AriaNN uses private equality testing proposed by \cite{fss2}.
The evaluators evaluate all possible paths before the special path (where the public value $a$ is equal to the private value $\alpha$). Then, they sum them up to obtain the final output of the comparison \cite{ryffel2021ariannlowinteractionprivacypreservingdeep}.

\begin{figure}[htp]
\centerline{\includegraphics[width=3in, trim={0cm 0.3cm 0cm 0.3cm}]{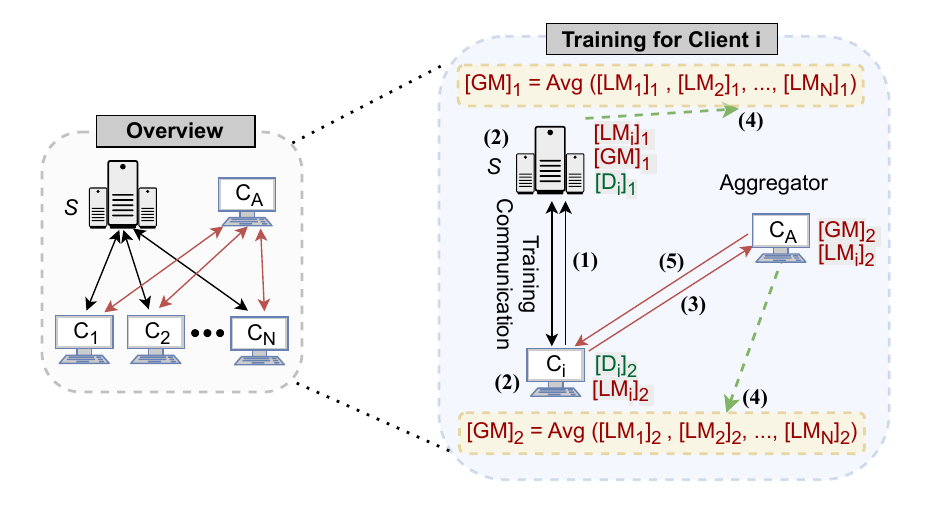}}
\caption{AriaNN-FL overview. (2) Each client collaboratively trains its local model with the server. The server holds a share of the model and the data, while the remaining shares are securely stored on the client side. (4) $C_A$, i.e., client aggregator, aggregates the client-side shares of the local models.}
\label{scheme2}
% \vspace{-0.3cm}
\end{figure}

AriaNN, then, extended their secure training framework to support federated learning, enabling privacy-preserving model training across several distributed clients.
The training process begins by secret-sharing the model parameters between the server and clients, each holds one share of the model. As illustrated in Fig.~\ref{scheme2}, the protocol proceeds as follows:  
(1) Each client secret-shares a portion of its local data with the server while retaining the other share locally.  
(2) Clients and the server then collaboratively train their respective shares of the local model using MPC for some training iterations.  
(3) After local training, one designated client (e.g., $C_A$) acts as the aggregator, collecting secret-shared updates from other clients.  
(4) The aggregator performs secure weighted averaging of these updates, ensuring that individual data and model privacy are preserved throughout. The server, in parallel, aggregates its corresponding shares of the updates. 
(5) The updated model share $[GM]_1$ is redistributed privately by the aggregator to all clients for the next round, while the server updates its copy $[GM]_0$ accordingly.

This process repeats for a fixed number of global epochs. At the end of training, the final model is reconstructed by combining the two secret shares: $GM = [GM]_0 + [GM]_1$
Since the model remains secret-shared throughout training, this design protects both data and model confidentiality while enabling collaborative learning.

% \vspace{-2mm}
\section{Fully Secure Federated Learning (FuSeFL)}

In this section, we present FuSeFL, an FL framework that holistically protects both data and model confidentiality while enabling scalable training. FuSeFL addresses key limitations of prior work, including privacy vulnerabilities and centralized bottlenecks. In existing systems, the server performs per-client training or coordination, limiting parallelism and creating a performance bottleneck. This becomes especially restrictive in settings where clients are data-rich. FuSeFL overcomes this by decentralizing training across client groups and assigning the server only lightweight computations for aggregation and group pairing, enabling secure and efficient scalability.

\subsection{Overview of FuSeFL}\label{intro}

\begin{figure}[htp]
\centerline{\includegraphics[width=3.4in, trim={0cm 0cm 0cm 0.5cm}]{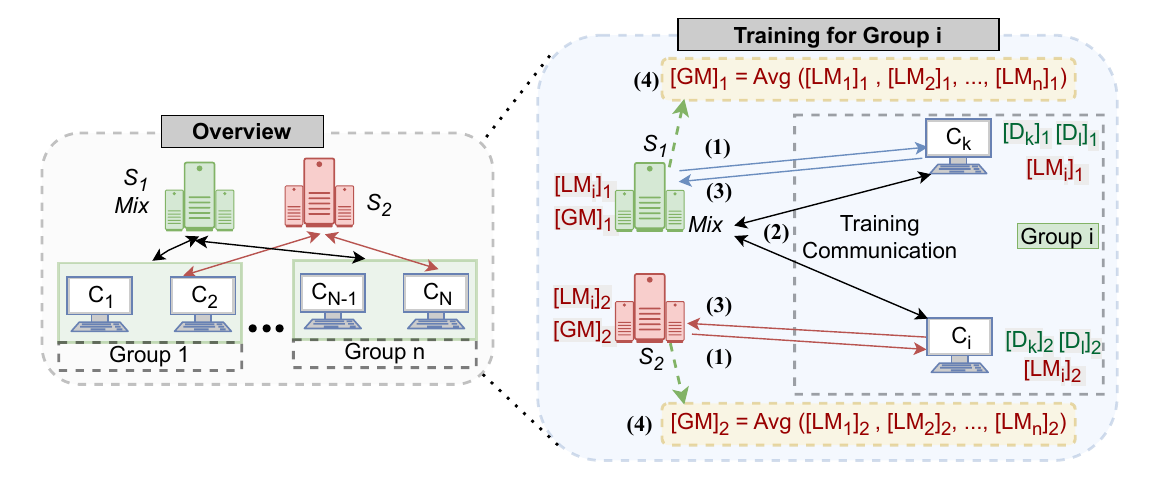}}
\caption{ Overview of FuSeFL. Clients shape groups of two. In each group, the data and model are secret shared and clients collaboratively and securely train the local model.}
\label{setting}
% \vspace{-0.4cm}
\end{figure}

To enable secure distributed training, FuSeFL leverages MPC to keep both data and model parameters secret-shared throughout. The protocol builds on AriaNN's MPC backend but rearchitects the system to support federated deployment, enhance scalability, and lower training time. 
To prevent collusion among computing parties, we assume that one server is trusted. This trusted server acts as a mix, coordinating the computing clients. All training-related data communication among groups is routed through it, ensuring that both senders and receivers remain anonymous.
In each training round, clients are anonymously grouped into pairs. Within each group, both the model and client data ($D_k$ and $D_L$) are secret-shared to ensure that no single party can access them in plaintext. As shown in Fig.~\ref{setting}, the workflow proceeds as follows:  

\begin{enumerate}[nosep,leftmargin=*]
\item \textbf{Model Distribution:} The global model is secret-shared between two servers ($S_1$, $S_2$), who then transmit the shares to the two clients ($C_k$, $C_l$) in each group. Specifically, $S_1$ sends $[GM]_1$ to $C_k$, and $S_2$ sends $[GM]_2$ to $C_l$.
\item \textbf{Secure Group Training:} Clients $C_k$ and $C_l$ collaboratively train a local model over their secret-shared data ($D_k$ and $D_L$) using MPC. In our case, we use AriaNN to perform secure training within a group (\S \ref{ariann}). The model itself and the data are maintained in a secret-shared form throughout training, ensuring confidentiality of both inputs and intermediate states. One share of the data remains local on each client to ensure that the data remains confidential and never leaves the client in its entirety. All data communication between pairs of clients in this stage is routed anonymously through the mix, $S_1$, which is the trusted server. $S_1$ pairs the clients using the Risk-Aware Dynamic Client Grouping algorithm and routes the training information accordingly. This additional communication layer is used to route, shuffle, and delay the packets in order to keep the pairs anonymous.
\item \textbf{Update Submission:} After completing local training for several epochs, each client returns their share of the updated local model to the corresponding server. $C_k$ sends $[LM]_1$ to $S_1$ and $C_l$ sends $[LM]_2$ to $S_2$.
\item \textbf{Secure Aggregation:} Servers compute a weighted average (FedAvg~\cite{mcmahan2023communicationefficientlearningdeepnetworks}) over the secret-shared model updates. Because updates are shared across two non-colluding servers, confidentiality is preserved even during aggregation.
\item \textbf{Iteration:} The aggregated global model is redistributed to client groups for the next training round. This process repeats for multiple global epochs.
\end{enumerate}

By structuring the training process around small, secure client groups and decoupling server-side training from aggregation, FuSeFL achieves a practical and scalable solution for secure FL.

\subsection{Risk-Aware Dynamic Client Grouping}

\newchanges{To reduce the likelihood of weakening anonymity through flow-matching attacks\cite{mixmatch} and backdoor channels that may lead to information leakage and collusion over time, FuSeFL employs a dynamic, adversary-aware grouping strategy in which client pairs are reshuffled every global round. This continual reconfiguration enhances robustness by minimizing repeated co-training relationships, thereby amplifying privacy through exposure diversity and repeated communication patterns. 
Each client pair is assigned a \emph{collusion risk score} $\rho(C_i, C_j) \in [0,1]$, reflecting potential anonymity weakening and structural correlations such as shared ownership, industry overlap, or regulatory proximity.} Given these scores, FuSeFL computes a set of disjoint client pairs $\mathcal{G} = \{(C_i, C_j)\}$ that minimize overall collusion exposure:
% \vspace{-2mm}
\begin{equation}
\begin{aligned}
\mathcal{G}^* = \arg\min_{\mathcal{G} \subset \binom{C}{2}} \; & \sum_{(C_i, C_j) \in \mathcal{G}} \rho(C_i, C_j) \\
\text{s.t.} \quad & \text{each } C_i \text{ appears exactly once.}
\end{aligned}
% \vspace{-2mm}
\end{equation}
This problem maps to a classical \emph{minimum-weight perfect matching} on a complete undirected graph, where clients are nodes and edge weights correspond to collusion risk scores. The exact solution can be obtained in $O(n^3)$ time using combinatorial optimization techniques such as the Hungarian algorithm~\cite{kolmogorov2020blossom}. However, for large-scale deployments, FuSeFL employs a more scalable approximation: it greedily constructs groupings by iteratively selecting the lowest-risk unpaired edge until all clients are matched. This greedy matching runs in $O(n^2 \log n)$ and has shown strong empirical performance for similar pairwise minimization tasks~\cite{price2023efficient}.

\textbf{Pipelined Execution:} FuSeFL performs group assignment computation entirely in parallel with training. Because group matching relies solely on static metadata and does not interact with model state or local data, assignments for round $t{+}1$ are computed concurrently while training for round $t$ is ongoing. As a result, this procedure incurs no additional latency in the training workflow.

\textbf{Establishing MPC Channels:} Once groups are selected, each pair $(C_i, C_j)$ anonymously initiates a secure MPC session using authenticated metadata exchange and ephemeral key agreement through the trusted server node. These sessions are short-lived and introduce negligible delay (typically under 30ms per pair~\cite{ryffel2021ariannlowinteractionprivacypreservingdeep}). Channels are secured via TLS or mutual-authenticated Diffie-Hellman, in alignment with enterprise-grade deployments.

This risk-aware, round-wise reshuffling reduces long-term collusion opportunities, ensures regulatory separation, and strengthens overall privacy guarantees in FL settings.

\begin{figure}[htp]
\centerline{\includegraphics[width=2.8in, trim={0.5cm 0.4cm 0.5cm 0.3cm}]{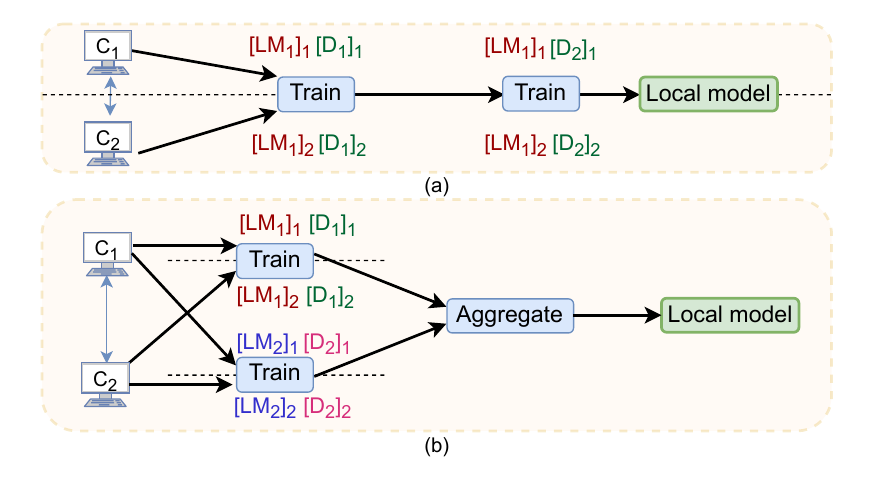}}
\caption{(a) FuSeFL-Serial trains the same model using the data of both clients in the group. (b) FuSeFL-Parallel trains two separate local models using the data of each client.}
\label{scheme3}
% \vspace{-3mm}
\end{figure} 

% \vspace{-1mm}
\subsection{Training Modes: Serial and Parallel}
To address the trade-offs between training accuracy, latency, and memory usage, FuSeFL introduces two intra-group training modes: \textit{FuSeFL-Serial} and \textit{FuSeFL-Parallel}, each offering distinct performance characteristics.

In \textit{FuSeFL-Serial} (Fig.~\ref{scheme3}(a)), each client group jointly trains a single model instance using the data from both clients, one after the other. First, the local model $LM_1$ is trained using client $C_1$’s data $D_1$ under MPC. Then, the same model is further trained using client $C_2$’s data $D_2$. This sequential training leverages the full dataset available to the group, leading to improved model accuracy. Since only one model instance is maintained, this mode is memory-efficient. However, it incurs higher training latency due to its serialized nature. 
% Algorithm~\ref{alg:one} provides implementation details.

% \SetKwComment{Comment}{/* }{ */}
% \vspace{-3mm}
% \begin{algorithm}
% % \footnotesize % or
% \scriptsize
% % for even smaller text
% \setstretch{0.05} % reduce line spacing, use \usepackage{setspace} in preamble
% \caption{FuSeFL-Serial}\label{alg:one}
% \textbf{Input:} $j \gets 2$ \\
% % \hspace{10mm}  Global model \([GM]_j\);\\ 
% % \hspace{10mm} Local models \([LM_{(i/2)}]_j\);\\ 
% % \hspace{10mm} Local datasets \( [D_i]_j \)\\

% \For{s in j}{ 
% Server s $\gets$  \([GM]_j\) \\ 
% }
% \For{each global epoch}{
    
%     \For{each group g (in parallel)}{
    
%         \For{each client c in g}{
%             \For{s in j}{ 
%                 Server s sends \([GM]_j\) to client c\\ 
%                 % Client c $\gets$  \([GM]_j\)
%             }
        
%             Client c $\gets$  \([D_c]_j\) \\
%             Client c $\gets$  \([D_{(j-c)}]_{(j-c)}\) \\
%              \([LM_g]_j\) $\gets$  \([GM]_j\)
%         }
%         \For{each local epoch}{
%         clients train \([LM_g]\) using \( [D_0] \)}
%         \For{each local epoch}{
%             clients train \([LM_g]\) using \( [D_1] \)
%         }
%         \For{each client c in g}{
%             \For{s in j}{ 
%                 client c sends \([LM_g]_j\) to server s
%             }
%         }
%     }
%      \([GM]_j\) = $AVG(\Sigma{[LM_g]_j)})$
% }
% \end{algorithm}
% \vspace{-3mm}

In contrast, \textit{FuSeFL-Parallel} (Fig.~\ref{scheme3}(b)) accelerates training by allowing both clients to train separate local models concurrently. The group maintains two model instances ($LM_1$ and $LM_2$), which are secret-shared among the clients. Each model is trained independently on $D_1$ and $D_2$ in parallel. After training, the clients locally aggregate the two updated models before securely transmitting the result to the servers. This mode reduces training latency and is better suited for time-sensitive deployments (evaluated in \S \ref{sec:computation}), at the cost of slightly higher memory usage and a modest trade-off in final model accuracy (evaluated in \S \ref{sec:accuracy}).
Details are outlined in Algorithm~\ref{alg:two}.
Both training modes fully preserve data and model confidentiality through MPC, ensuring that no single client or server can reconstruct sensitive information.

% Table~\ref{tradeoffs-table} summarizes the trade-offs between the two modes.

% \begin{table}[htp]
% % \vspace{2mm}
% 	\centering
% 	\caption{FuSeFL-Serial vs. FuSeFL-Parallel --- performance and resource trade-offs.}
%     \vspace{-2mm}
% 	\scalebox{0.70}{
% 	\begin{tabular}{>{\centering\arraybackslash}p{1.0in}>
%     {\centering\arraybackslash}p{1.5in}>
%     {\centering\arraybackslash}p{1.2in}}

% 		\toprule
%         \midrule
%         \textbf{Aspect} & \textbf{FuSeFL-Serial} & \textbf{FuSeFL-Parallel} \\
% 		\midrule

%         Training Latency & Higher & Lower \vspace{1mm} \\
%         \midrule
%         Model Accuracy & Slightly higher & Slightly lower \vspace{1mm} \\
%         \midrule
%         Memory Usage & Lower  & Higher \vspace{1mm} \\
%         \midrule
%         Best Use Case & Accuracy- or memory- sensitive applications & Latency-sensitive applications \vspace{1mm} \\

% 		\midrule
% 		\bottomrule
% 	\end{tabular}}
    
% 	\label{tradeoffs-table}
%     \captionsetup{font=small}
%     \vspace{-5mm}
% \end{table}

\SetKwComment{Comment}{/* }{ */}
% \vspace{-2mm}
\begin{algorithm}
% \footnotesize % or 
\scriptsize 
% for even smaller text
\setstretch{0.05} % reduce line spacing, use \usepackage{setspace} in preamble
\caption{FuSeFL-Parallel}\label{alg:two}

\For{each global epoch}{
    
    \For{each group g (in parallel)}{
    
        \For{each client c in g}{
            \For{s in j}{ 
                Server s sends \([GM]_j\) to client c\\ 
                % Client c $\gets$  \([GM]_j\)
            }
        
            Client c $\gets$  \([D_c]_j\) \\
            Client c $\gets$  \([D_{(j-c)}]_{(j-c)}\) \\
             \([LM_c]_j\) $\gets$  \([GM]_j\)
        }
        \For{each local epoch}{
        clients train \([LM_c]\) using \( [D_0] \)}
        \For{each local epoch}{
            clients train \([LM_g]\) using \( [D_1] \)
        }
        $[LM_g] = Aggregate([LM_{2n}], [LM_{2n+1}])$
        
    }
     \([GM]\) = $AVG(\Sigma{[LM_g])})$
}
\end{algorithm}
\vspace{-3mm}

\subsection{Communication Complexity} \label{describe:comm}
Secure FL systems incur communication overhead in two channels: server-to/from-client (S2C/C2S) and client-to-client (C2C).

\textbf{Server-to-Client / Client-to-Server (S2C/C2S). } 
In AriaNN-FL, the server actively participates in secure training by running MPC protocols jointly with each client across every local epoch. This 
% results in $\mathcal{O}(n \cdot x \cdot e)$ communication per global round and 
introduces a severe server-side computational bottleneck as the number of clients grows.  
In contrast, FuSeFL decouples training from the server and restricts its role to model aggregation, client pairing, and data routing. Clients in each group exchange data through a trusted server, and at the end of each round, they send their secret-shared updates to two non-colluding servers. The overall direct S2C/C2S complexity for aggregation and distribution is reduced to $\mathcal{O}(n \cdot x)$, assuming a model size of $x$ and $n$ total clients. This significantly improves scalability and enables parallel execution.
The data communication required to distribute clients’ data, however, increases from $\mathcal{O}(n \cdot d)$ to $\mathcal{O}(n \cdot (2 \cdot d))$, assuming that the redirection delay at the server is negligible, where $d$ is the data size of each client and $n$ is the total number of clients.

\textbf{Client-to-Client (C2C). }
AriaNN-FL performs client-to-client communication during aggregation and distribution, where each client sends its model share to a designated aggregator for reconstruction and gets back the updated global model, which incurs $\mathcal{O}(2 \cdot n \cdot x)$ communication overhead, assuming a model size of $x$ and $n$ total clients.
In contrast, FuSeFL confines secure training to fixed client pairs using MPC. Each pair exchanges secret shares of data and model parameters through the trusted server. This increases the per-group C2C overhead to achieve a higher level of security, which is the non-colluding clients. constant and independent of the total number of clients.
By restructuring the communication pattern to exploit pairwise training and two-server aggregation, FuSeFL substantially reduces the training time while increasing the communication overhead (evaluated in \S \ref{sec:computation} \S and \ref{sec:communication}). This architectural change eliminates server bottlenecks and enables system-wide parallelism without weakening security guarantees.

% \begin{table}[ht]
% \centering
% \caption{Communication Complexity of AriaNN-FL vs. FuSeFL}
% \label{tab:comm-complexity}

% \scalebox{0.70}{
% \begin{tabular}{
% >{\centering\arraybackslash}p{1.5in}
% >{\centering\arraybackslash}p{1.3in}
% >{\centering\arraybackslash}p{1.3in}}
% \toprule
% \midrule
% \textbf{Communication Type} & \textbf{AriaNN-FL} & \textbf{FuSeFL} \\
% \midrule
% S2C / C2S & $\mathcal{O}(n \cdot x \cdot e)$ & $\mathcal{O}\left(\frac{n}{2} \cdot x \cdot e\right)$ \vspace{1mm} \\
% C2C & $\mathcal{O}(n \cdot x)$ & $\mathcal{O}(x)$ (per group) \vspace{1mm} \\
% \midrule
% \bottomrule
% \end{tabular}}
% \captionsetup{font=small}
% \vspace{1mm}
% \caption*{\small $x$: model size; $n$: number of clients; $e$: local epochs.}
% \label{tab:comm-complexity}
% \end{table}

% \vspace{-1mm}
\subsection{The Memory Complexity}\label{memexplained}

Table~\ref{memory2} summarizes the number of model instances maintained by different entities under AriaNN-FL and FuSeFL (in both Serial and Parallel modes). In AriaNN-FL, the server is involved in both secure training and aggregation, requiring it to maintain $n$ local model instances—one per client—plus the global model. Similarly, the designated aggregator client stores $n$ model shares for aggregation, resulting in significant memory overhead concentrated on central parties.
In contrast, FuSeFL distributes the training workload across client pairs, thereby decentralizing memory usage. Each of the two aggregation servers stores only $(n/2) + 1$ model instances, cutting the server-side memory requirement by nearly 50\%. On the client side, FuSeFL-Serial maintains a single model per group, yielding comparable memory usage to AriaNN-FL but with distributed storage. FuSeFL-Parallel trades memory for speed: each client trains an independent local model copy, increasing client-side memory usage to $3n$ total model instances across all clients. This redundancy enables parallel execution and accelerates training without incurring additional server load.

\subsection{Security Guarantees}
FuSeFL provides strong security guarantees for both data and model confidentiality in a distributed, partially untrusted environment. These guarantees are realized through a combination of secret sharing, MPC, and adversary-aware system design, as defined in our threat model (Section~\ref{thretmod}). Below, we detail the key properties and the cryptographic principles underpinning them.

\begin{table}[H]
% \vspace{-1mm}
	\centering
 % \color{blue}
	\caption{Number of model instances stored on various parties. }
    % \vspace{-2.7mm}
	\scalebox{0.70}{
	\begin{tabular}{>{\centering\arraybackslash}p{1.2in}>{\centering\arraybackslash}p{0.8in}>
    {\centering\arraybackslash}p{0.9in}>
    {\centering\arraybackslash}p{1.0in}}
    
		\toprule
        \midrule
  &AriaNN-FL&FuSeFL-Serial&FuSeFL-Parallel\\
		\midrule
  
    {Server1}&n + 1&$(n\div2) + 1$&$(n\div2) + 1$\vspace{1mm}\\
    {Server2}&N/A& $(n\div2) + 1$& $(n\div2) + 1$ \vspace{1mm}\\
    {Clients}&n& n& 3n \vspace{1mm}\\
    {Client Aggregator}&n + 1& N/A & N/A \vspace{1mm}\\
    
\midrule
		\bottomrule
	\end{tabular}}
    
	\label{memory2}
    \captionsetup{ font=small}
    % \vspace{-3mm}
    %\caption*{Note: N/A means there is no such party in that scheme. For example, AriaNN-FL is based on one server, and no second server is involved.}
\end{table}

\begin{itemize}[nosep,leftmargin=1.5em]

\item \textbf{Confidentiality of Local Data, Model Parameters, and Updates:} 
Each client retains the confidentiality of its data, model parameters, and model updates throughout the training process. FuSeFL employs additive secret sharing to ensure that no single party has access to the plaintext of any sensitive information. All values exchanged during training are masked using independently sampled random shares that are statistically indistinguishable from uniform noise. Secure evaluation of both linear and nonlinear operations is achieved using function sharing techniques as in AriaNN~\cite{ryffel2021ariannlowinteractionprivacypreservingdeep}. Since neither client nor server ever observes unmasked values, the protocol provides semantic security against honest-but-curious adversaries.

\item \textbf{Correctness of Secure Training:} 
FuSeFL builds directly on the AriaNN protocol, which guarantees the correctness of MPC-based training among two clients. Each party's local execution follows the prescribed sequence of arithmetic and Boolean operations in the secure computation circuit. The resulting model update is functionally equivalent to that of plaintext training. For formal analysis, we refer to the proofs provided in~\cite{ryffel2021ariannlowinteractionprivacypreservingdeep}.

\item \textbf{Correctness of Aggregation:} 
After each group completes training, clients submit secret shares of the locally updated model to two non-colluding aggregation servers. Each server independently performs FedAvg-style aggregation~\cite{mcmahan2023communicationefficientlearningdeepnetworks} over the received shares. Since weighted averaging is a linear operation, it naturally composes with additive secret sharing. Thus, the final global model obtained after reconstruction matches the federated average of all local updates. The correctness of this aggregation step is inherited from prior secure aggregation frameworks~\cite{secureml, aby, ABY3}.

\item \textbf{Protection Against Curious Clients:} 
Clients never access the raw data or model parameters of their partner. Instead, they hold one share of their partner's input and model, making reconstruction infeasible under an honest-but-curious assumption. This ensures that a semi-honest client cannot learn the counterpart’s training data or model parameters. 
% Note that FuSeFL does not assume malicious collusion between group members; however, privacy guarantees degrade proportionally with the number of colluding parties.
FuSeFL’s security model assumes that at least one aggregation server is trusted and non-colluding, ensuring that no collusion occurs between the two servers. Inspired by the concept of mix networks \cite{mixnet1,mixnet2}, this trusted server is additionally responsible for anonymously pairing clients during each training round. By routing all inter-client communication through the trusted server, FuSeFL prevents clients from directly identifying or interacting with their training partners, thereby mitigating the risk of collusion and information leakage.

\end{itemize}

% \vspace{-2mm}
\subsection{ Flexibility and Generalization}

FuSeFL is designed to be modular with respect to the underlying secure computation protocol. While our current implementation leverages the AriaNN scheme for secure collaborative training within client groups, FuSeFL's architecture is not bound to this specific backend. The abstraction of secure local training allows FuSeFL to seamlessly generalize to a wider class of MPC schemes.
For instance, adopting a three-party MPC protocol such as SecureNN~\cite{securenn} or Falcon~\cite{falcon} enables FuSeFL to extend its grouping strategy to triads of clients. In such a configuration, each client secret-shares its data and model with the remaining clients in the group, maintaining one share locally. Training proceeds securely over these shared inputs without revealing sensitive data to any single party. The model itself must also be secret-shared in alignment with the MPC protocol’s trust assumptions.
To maintain end-to-end confidentiality, the aggregation infrastructure must scale accordingly. Specifically, FuSeFL requires one aggregation server per client in the group to receive and process the respective secret shares. 
This flexible design allows FuSeFL to support a broad range of MPC backends with different trust and performance trade-offs, making it suitable for deployment in diverse FL environments where regulatory, computational, or adversarial constraints may vary.

% \vspace{-0.2cm}
\section{Evaluation Methodology}\label{sec:eval-method}

To assess the performance and accuracy of FuSeFL, we adopt the same experimental setup and neural network architectures used in the AriaNN benchmark~\cite{ryffel2021ariannlowinteractionprivacypreservingdeep}. Specifically, we evaluate FuSeFL against our baselines using three network topologies:

\begin{itemize}[nosep,leftmargin=1.5em]
\item \textbf{Network-1:} A multilayer perceptron composed of three fully connected layers, each followed by ReLU activation functions.
\item \textbf{Network-2 and LeNet:} A convolutional neural network (CNN) with two convolutional layers (with ReLU and max pooling) followed by two fully connected layers.
% \item \textbf{LeNet:} A classic CNN architecture similar to Network-2 but with different layer dimensions and configurations.
\end{itemize}

MNIST dataset~\cite{mnist} is used to train these networks. This dataset is uniformly partitioned among all participating clients. Each client receives an equal share of the data and contributes a single vote to the federated average (i.e., FedAvg weights are uniform across clients). Experiments are conducted on a machine equipped with an Intel(R) Xeon(R) W-2123 CPU (3.60 GHz, 4 cores). Our FuSeFL implementation is built on top of AriaNN's implementation~\cite{ryffel2021ariannlowinteractionprivacypreservingdeep}, which performs secure training using two workers while keeping the data and model private. In our setup, both workers and the coordinating server run as separate processes on the same physical machine, communicating over a local network (LAN). In FuSeFL, the AriaNN training protocol is applied within each client group to collaboratively train the local model.

% \vspace{-2mm}
\section{Evaluation Result} \label{sec:eval}
This section evaluates FuSeFL across multiple dimensions, including model accuracy, communication overhead, memory footprint, and training time. We use three baselines for comparison: (1) AriaNN, the original one-server, one-client secure training protocol proposed by Ryffel et al.~\cite{ryffel2021ariannlowinteractionprivacypreservingdeep}, (2) AriaNN-FL, the FL extension of AriaNN, and (3) WW-FL, a fully offloaded secure FL scheme \cite{wwfl}.

% \vspace{-2mm}
\subsection{Server-Side Scalability Analysis}\label{sec:computation}

\begin{figure}[htp]
\centerline{\includegraphics[width=3.4in, trim={2cm 3.5cm 15cm 2.9cm}]{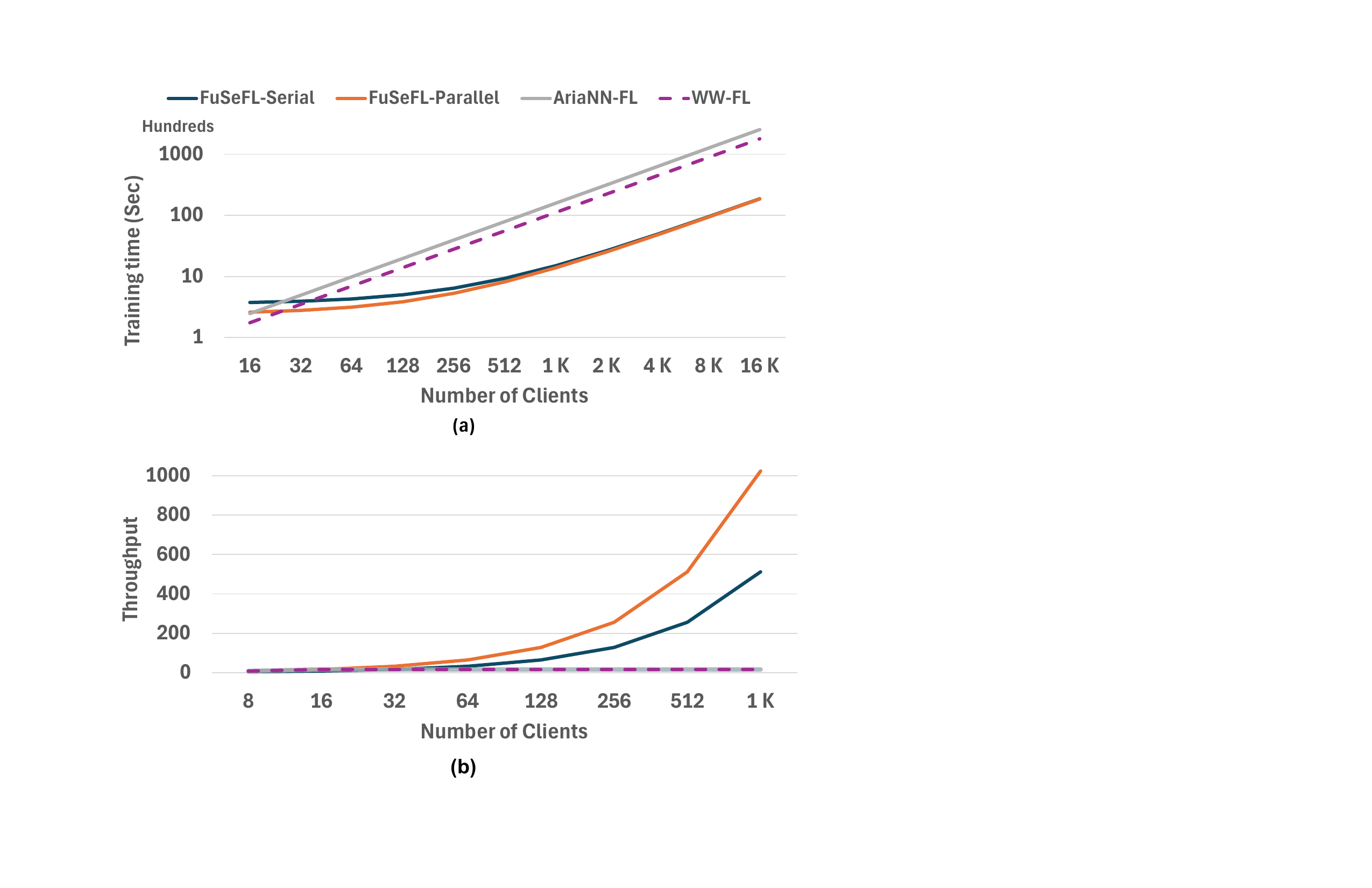}}
\caption{(a) Training throughput per round. FuSeFL's throughput is not limited by the server's computational capabilities. (b)Training time per epoch across different schemes as the number of clients increases.}
% thereby minimizing the number of model instances the server must store.}
\label{timing}
% \vspace{-0.1cm}
\end{figure}

To assess FuSeFL’s server-side scalability, we evaluate its ability to handle an increasing number of clients under fixed hardware resources, focusing on the server's computational overhead for training and aggregation.
As illustrated in Fig.~\ref{timing}(a), AriaNN-FL's training time per epoch grows significantly with the number of clients, due to its centralized training coordination. For example, with 16 K clients and a dataset of size 46 MB per client, FuSeFL-Serial achieves up to a $13.50\times$ speedup over AriaNN-FL, while FuSeFL-Parallel achieves up to $13.60\times$. The bottleneck in AriaNN-FL arises from limited parallelism: if a server has 16 compute threads, it must process clients in sequential batches, e.g., training 64 clients in four rounds of 16. This serialization leads to latency that scales linearly with client count, severely limiting scalability.
FuSeFL avoids this bottleneck by shifting training to client pairs. Servers are only responsible for secure aggregation and routing, not training. Since client-group training occurs in parallel and independently of server compute capacity, FuSeFL's training time increases steadily as the number of clients increases. This design enables scalable, low-latency training suitable for large-scale cross-silo deployments.

To compare WW-FL~\cite{wwfl} with AriaNN and FuSeFL, we configure this system to use AriaNN as its underlying MPC protocol and evaluate them under identical training parameters. Notably, FuSeFL performs local training without relying on third-party computation servers, whereas WW-FL requires external servers within each cluster to train secret-shared client data.
As illustrated in Fig.~\ref{timing}(a), WW-FL’s training time increases linearly with the number of clients. This is because more clients are assigned to each server during training. Although the communication overhead is distributed across the network, computation can become the bottleneck as the number of clients grows. For example, with 16 K clients, a dataset size of 46 MB per client, and 16 server clusters, FuSeFL-Serial and FuSeFL-Parallel achieve up to a $9.50\times$ and $9.60\times$ speedup over WW-FL, respectively.

\begin{figure*}[!t]
\centerline{\includegraphics[width=0.85\textwidth, trim={0cm 0cm 0cm 0cm}]{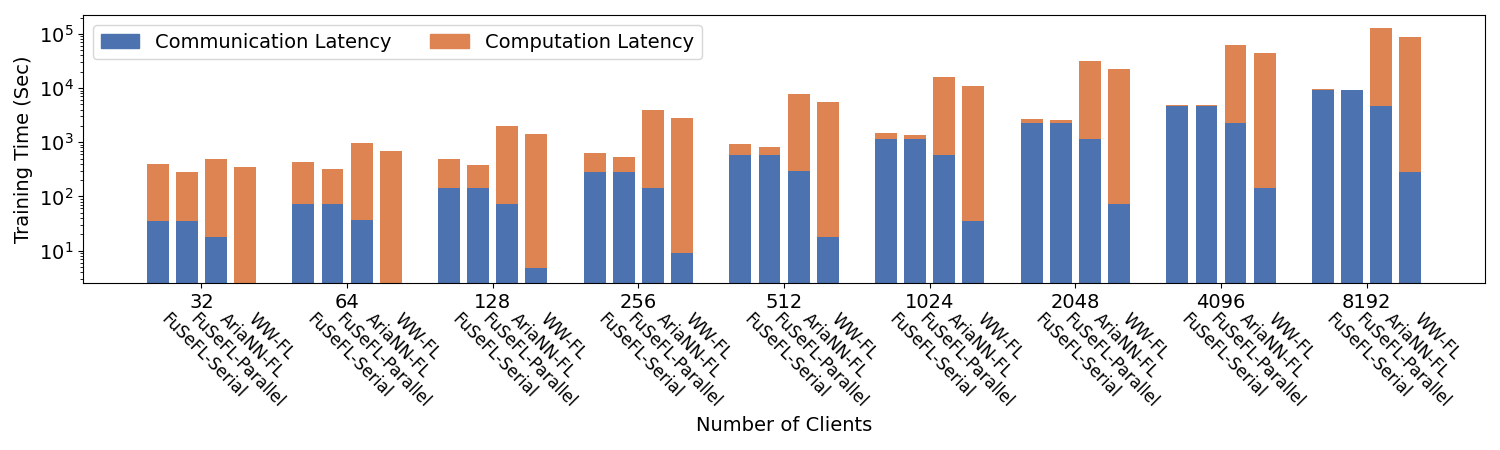}}
\caption{Detailed training time per epoch across different schemes. In FuSeFL, computation time is negligible when increasing the number of clients. }
\label{timingdetail}
% \vspace{-0.3cm}
\end{figure*}

As shown in Fig. \ref{timingdetail}, FuSeFL’s end-to-end training time becomes increasingly dominated by communication latency as the number of clients increases, while computation latency remains comparatively small. This is because computation is distributed among clients rather than a limited number of servers. In contrast, AriaNN-FL and WW-FL incur substantially higher total training times; although their communication overhead is lower than FuSeFL’s, their computation overhead becomes the bottleneck at larger client counts.
Adding a trusted server as a routing hop between the clients results in an increase in the communication time in FuSeFL, which is approximately $2\times$ and $32\times$ higher than that of AriaNN and WW-FL, respectively. However, by reducing the computation time, FuSeFL can achieve a better overall training time.

To further highlight the server-side bottleneck, Fig.~\ref{timing}(b) compares the training throughput, measured as the number of clients trained in parallel, for AriaNN-FL, WW-FL, and FuSeFL. In both AriaNN-FL and WW-FL, throughput initially increases with client count but quickly saturates once the number of clients exceeds the server’s available CPU cores (in AriaNN-FL) or the number of server clusters (in WW-FL). This plateau reflects limited server parallelism and growing serialization overhead as more clients participate. In contrast, FuSeFL decouples training from server compute constraints. Since training is offloaded to client pairs, throughput scales linearly with the number of clients, demonstrating FuSeFL’s superior scalability under high client loads.

\begin{figure*}[!t]
\centerline{\includegraphics[width=0.70\textwidth, trim={2cm 13cm 4cm 3.8cm}]{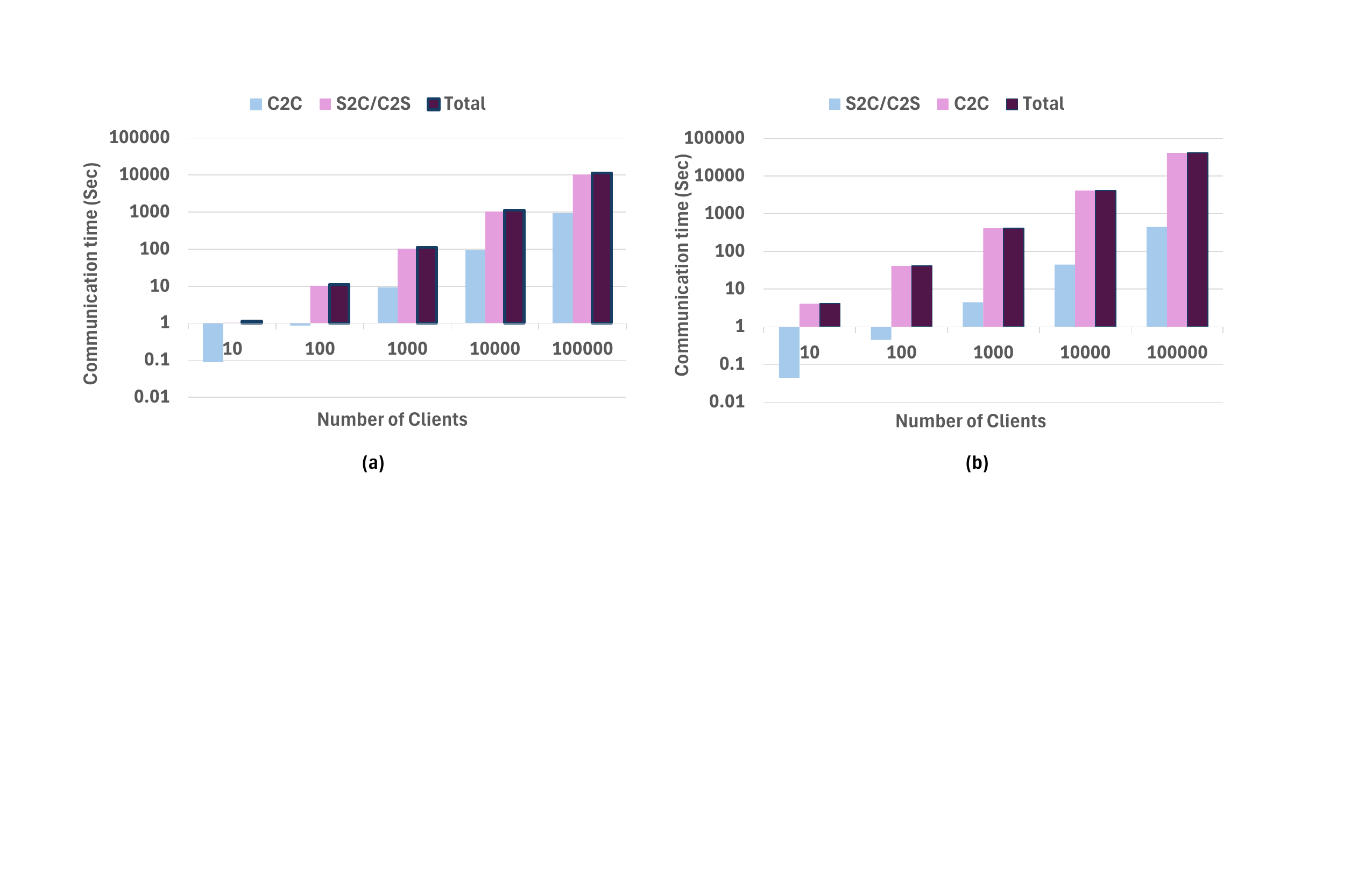}}
\caption{Communication Overhead in FuSeFL vs. AriaNN-FL.
%Communication overhead of FuSeFL compared to AriaNN-FL. By distributing the training across clients, the required bandwidth is spread over the network, resulting in significantly lower client-to-client (C2C) communication overhead.
}
\label{communication}
% \vspace{-0.2cm}
\end{figure*} 

% \vspace{-0.2cm}
\subsection{Communication Latency at Scale}\label{sec:communication}

We measure the total communication time for one training epoch of Network1 under AriaNN-FL and FuSeFL, assuming each client holds a dataset of 46 MB. We break down the overhead into two channels described on \S\ref{describe:comm}: C2C and S2C/C2S, as shown in Fig.~\ref{communication}.

FuSeFL’s communication overhead exceeds that of AriaNN-FL due to the need to support non-colluding parties. Inspired by the idea of mixnets, we propose using a trusted node (one of the servers in our case) as a bridge to connect two computing parties anonymously. As shown in Fig.~\ref{communication}(b), in this design, C2C communication is routed through the trusted server. As a result, this overhead increases linearly with the number of clients.
S2C/C2S communication in FuSeFL, which only involves transmitting the final model shares to the servers for secure aggregation, also scales linearly but at half the rate of AriaNN-FL. This is because only one model per group is uploaded per round, cutting the communication volume in half. 
It is true that FuSeFL’s communication latency is twice that of AriaNN-FL, but the benefits FuSeFL gains by distributing the training cancel out this drawback, specifically, without losing any security guarantees.
% For instance, at 10,000 clients, FuSeFL's total communication time is only 45.21 seconds compared to 1,123 seconds in AriaNN-FL, yielding a 25× reduction.
This efficiency gain highlights FuSeFL’s ability to scale securely and efficiently across thousands of clients.
% , where communication bandwidth and latency are critical deployment constraints.
Although FuSeFL employs a mixnet-style trusted routing layer to anonymize communication between clients, the additional delay introduced by this mechanism is negligible. For example, in our experiments with 256 clients, the per-pair routing and shuffling delay was approximately \textbf{2\%} of the total communication time and did not affect overall training throughput. Thus, the mixnet integration in FuSeFL provides robust privacy guarantees with minimal performance cost.

% \vspace{-0.2cm}
\subsection{Memory Overhead}\label{sec:memory}
\begin{figure}[htp]
\centerline{\includegraphics[width=2.5in, trim={3cm 7cm 4cm 3cm}]{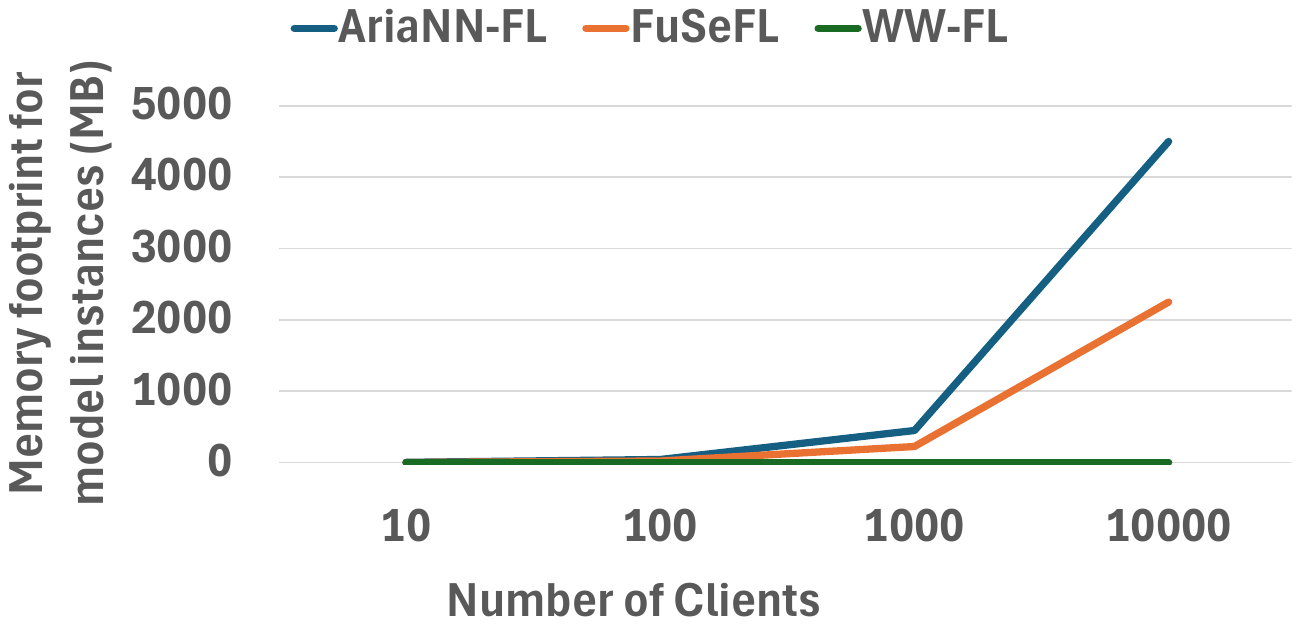}}
\caption{Memory footprint of model instances at the server.
FuSeFL reduces server-side memory usage by 50\% compared to AriaNN-FL by offloading training to client groups.}
% thereby minimizing the number of model instances the server must store.}
\label{memory}
% \vspace{-0.3cm}
\end{figure}

Aggregator nodes in federated learning must store multiple instances of local model parameters until all client updates are received for weighted aggregation at the end of each epoch~\cite{flagger}. This requirement leads to considerable memory overhead, especially as the number of clients increases.
Figure~\ref{memory} shows the server-side memory usage for storing model instances for Network1. In conventional FedAvg, the server cannot incrementally aggregate updates and must retain all client updates simultaneously, causing memory consumption to scale linearly with the number of participants. In WW-FL, however, the memory usage on the server side remains steady as the number of clients increases. This amount is limited by the number of server clusters used for training. On the server, we need to store one instance of the model per server cluster.

FuSeFL improves this by organizing training into client pairs, each producing a single aggregated model update per round. As a result, the number of model instances stored on the server is effectively halved. %As illustrated in Figure~\ref{memory} and described in Section~\ref{memexplained}, FuSeFL reduces aggregator memory usage by nearly 50\%.
For instance, with 128 clients, AriaNN-FL stores 128 local models plus the global model. FuSeFL reduces this to 64 local models while preserving model quality and reducing memory load. This improvement is especially beneficial in cross-silo deployments, where the number of participating institutions can be large and model sizes (e.g., VGG-16 at 528MB in FP32) significantly increase server storage demands.

\begin{figure*}[!t]
\centerline{\includegraphics[width=0.72\textwidth, trim={0cm 0.5cm 0cm 1.5cm}]{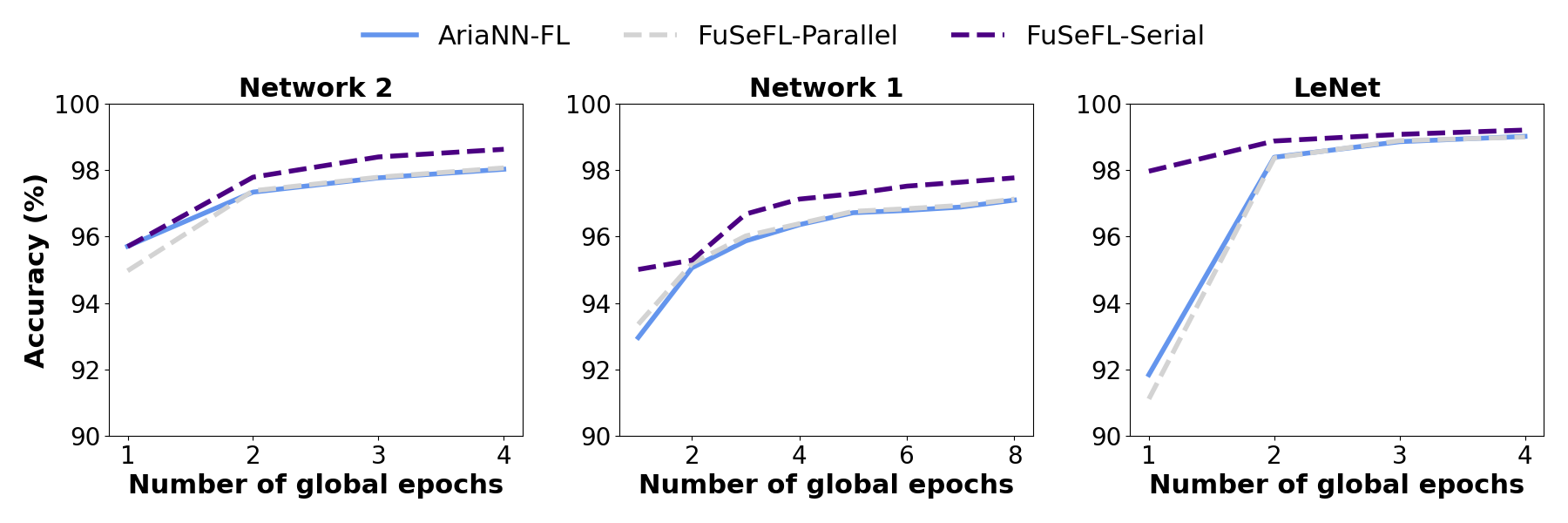}}
\caption{Accuracy of different FL schemes for various global epochs and 4 local epochs.}
\label{acc3}
% \vspace{-0.2cm}
\end{figure*}

\begin{table}[htp]
% \vspace{-1mm}
	\centering
 % \color{blue}
	\caption{Accuracy comparison of different schemes}
    % \vspace{-2.5mm}
	\scalebox{0.68}{
	\begin{tabular}{>{\centering\arraybackslash}p{1.2in}>{\centering\arraybackslash}p{0.6in}>
    {\centering\arraybackslash}p{0.9in}>
    {\centering\arraybackslash}p{0.9in}}
 % >{\centering\arraybackslash}p{0.9in}cccc}

		\toprule
        \midrule
  Scheme&Epochs*&Network&Accuracy(\%)\\
		\midrule
  
    {AriaNN\cite{ryffel2021ariannlowinteractionprivacypreservingdeep}}&15&Network1&\textbf{98.00}\vspace{1mm}\\
    {AriaNN-FL\cite{ryffel2021ariannlowinteractionprivacypreservingdeep}}&4/4& Network1& 96.40 \vspace{1mm}\\
    {FuSeFL-Parallel}&4/4& Network1& 96.40 \vspace{1mm}\\
    {FuSeFL-Serial}&4/4& Network1& 97.20 \vspace{1mm}\\
    
    \midrule
    {AriaNN\cite{ryffel2021ariannlowinteractionprivacypreservingdeep}}&10&Network2&\textbf{98.40}\vspace{1mm}\\
    {AriaNN-FL\cite{ryffel2021ariannlowinteractionprivacypreservingdeep}}&3/4& Network2& 97.80 \vspace{1mm}\\
    {FuSeFL-Parallel}&3/4& Network2& 97.80 \vspace{1mm}\\
    {FuSeFL-Serial}&3/4& Network2& \textbf{98.40} \vspace{1mm}\\

    \midrule
    {AriaNN\cite{ryffel2021ariannlowinteractionprivacypreservingdeep}}&10&LeNet&\textbf{99.20}\vspace{1mm}\\
    {AriaNN-FL\cite{ryffel2021ariannlowinteractionprivacypreservingdeep}}&4/4& LeNet& 99.00 \vspace{1mm}\\
    {FuSeFL-Parallel}&4/4& LeNet&99.00  \vspace{1mm}\\
    {FuSeFL-Serial}&4/4& LeNet& \textbf{99.20} \vspace{1mm}\\
    
    \midrule
		\bottomrule
	\end{tabular}}
    
	\label{accuracy}
    \captionsetup{ font=small}
    \vspace{1mm}
    \caption*{Note: (*) For the FL-based models, 4/4 means the model is trained for four global iterations and four local epochs (global/local).}
    % \vspace{-8.5mm}
\end{table}

% \vspace{-3mm}
\subsection{Accuracy Evaluation}\label{sec:accuracy}
Table~\ref{accuracy} presents a comparison of test accuracy for FuSeFL and AriaNN when training Network1, Network2, and LeNet on the MNIST dataset, evenly partitioned across 16 clients. AriaNN achieves 98\% accuracy on Network1 after 15 epochs of centralized secure training. However, extending AriaNN to a federated setting introduces accuracy degradation: with 4 global and 4 local epochs, AriaNN-FL and FuSeFL-Parallel reach 96.40\%, while FuSeFL-Serial achieves 97.20\%.
This accuracy gain in FuSeFL-Serial stems from its sequential intra-group training: each group first trains on one client’s data, then continues training on the second client’s data before aggregation. Aggregation is performed per group rather than per client, reducing data fragmentation.
Importantly, FuSeFL-Serial achieves comparable accuracy $1.43\times$ faster than AriaNN-FL and $6.85\times$ faster than AriaNN. For Network2 and LeNet, FuSeFL-Serial matches AriaNN’s accuracy (98.40\% and 99.20\%, respectively), demonstrating that FuSeFL delivers the same learning quality, while offering substantial speedups due to its decentralized design. AriaNN-FL and FuSeFL-Parallel yield slightly lower accuracy for Network2 and LeNet, reaching 97.80\% and 99.00\%, respectively.

\textbf{Accuracy Convergence Over Global Epochs:} 
Fig. \ref{acc3} shows how model accuracy evolves with increasing global training epochs for three networks (Network 1, Network 2, and LeNet). Across all models, both variations of FuSeFL (Serial and Parallel) demonstrate rapid convergence and maintain consistently high accuracy, closely tracking or outperforming AriaNN-FL. The FuSeFL-Serial variant offers a slight advantage in accuracy due to richer local training from sequential data access.

% \vspace{-2mm}
\subsection{Robustness Under Threat Model}

This section analyzes the robustness of FuSeFL under the adversarial threat model described in Section~\ref{thretmod}. The primary security goals are to protect the confidentiality of user data, model parameters, and intermediate updates in the presence of semi-honest adversaries. We consider two prominent threat vectors: unauthorized observation and model theft.

% \vspace{1mm}
\noindent\textbf{Protection Against Unauthorized Observation.} 
Adversaries may attempt to extract sensitive information by intercepting communication between clients or by observing model parameters during training. In FuSeFL, all private values, including client data, model parameters, and gradients, are secret-shared using additive masking. These shares appear statistically indistinguishable from random noise, preventing inference even in the case of eavesdropping.
To securely perform training, FuSeFL employs the AriaNN protocol~\cite{ryffel2021ariannlowinteractionprivacypreservingdeep}, which guarantees confidentiality through function sharing and additive masking. During secure training, no single party observes plaintext data or model parameters. Additionally, each client retains one share of its data locally and never reveals its complete input to its training partner, preserving local privacy. These guarantees are backed by established cryptographic protocols validated in prior literature~\cite{secureml,fss1,fss2,aby,ABY3}.

% \vspace{1mm}
\noindent\textbf{Protection Against Model Theft.} 
Model confidentiality is a critical requirement in federated settings, particularly when the model represents proprietary intellectual property. In FuSeFL, the global model is secret-shared between two non-colluding servers (at least one of which is trusted) and then transmitted to clients as encrypted shares. Each client receives only one share and cannot reconstruct the model independently. Even if an adversary intercepts a share during transmission, it reveals no meaningful information. In addition, the design of FuSeFL is backed up by the idea of mix networks, where the trusted server acts as a mix node to ensure that clients are grouped anonymously and cannot collude.
Furthermore, aggregation servers do not have access to the full model updates; they operate only on shares during the FedAvg procedure. These protections collectively ensure that no party, client or server, can exfiltrate the full model under the semi-honest assumption. FuSeFL thereby mitigates the risk of model theft during both training and communication phases.

% \vspace{-2mm}
\section{Related Work}

To protect privacy in FL, several cryptographic and statistical methods have been proposed. HE allows computation over encrypted data but is computationally intensive \cite{8241854, 10.1145/3133956.3133982}. DP introduces calibrated noise to gradients or updates to provide formal privacy guarantees \cite{McMahan2017, mcmahan2018}, but this often comes at the cost of reduced model accuracy. MPC, in contrast, offers strong privacy guarantees without degrading accuracy, though it typically incurs high communication costs.
MPC-based approaches have gained traction in secure FL. SecureML \cite{secureml}, SecureNN \cite{securenn}, Falcon \cite{falcon}, and AriaNN \cite{ryffel2021ariannlowinteractionprivacypreservingdeep} propose secure training protocols using various cryptographic primitives, including additive secret sharing \cite{ABY3, aby}, function secret sharing \cite{fss1, fss2}, and Yao's circuits \cite{yao}. These works typically offload both data and models to trusted third-party servers for secure computation. However, they assume trust in infrastructure and require clients to fully transmit their data shares, increasing bandwidth demands and collusion risks.
WW-FL \cite{wwfl} extends this idea by introducing a cluster-based MPC training framework. Clients are grouped into clusters and offload secret-shared data to server pools, which perform collaborative training. While this improves server-side load distribution, it still assumes trusted execution on third-party servers and leads to high communication overhead as the number of clients or cluster sizes increase. Additionally, WW-FL relies on a client-side trusted zone for safely outsourcing data, which may not be feasible in practical deployments.

Other works attempt to establish model ownership through watermarking \cite{waffle, liang2023fedcipfederatedclientintellectual}. Backdoor watermarking uses a trigger dataset, while parameter watermarking embeds ownership directly in model weights. While effective for post-training verification, these methods do not provide confidentiality during training and are orthogonal to secure computation.

\section{Conclusion}

In this work, we introduced FuSeFL, a fully secure and scalable federated learning framework tailored for cross-silo environments. By decentralizing training across client pairs using lightweight MPC and assigning a minimal aggregation role to the servers, FuSeFL eliminates central bottlenecks and ensures complete confidentiality of client data, model parameters, and training updates. Our design not only mitigates key inference threats, including model inversion and gradient leakage, but also significantly reduces training time and server resource demands. Our evaluations show that FuSeFL achieves performance improvements over existing secure FL solutions, establishing it as a practical and robust alternative for secure collaborative learning in regulated domains.

\bibliographystyle{IEEEtran}
\bibliography{IEEEabrv.bib, references.bib}

\end{document}